\documentclass[useAMS,usenatbib]{mn2e}
\usepackage{epsfig,lscape} 
\usepackage{natbib}
\usepackage{amsmath}
\usepackage{graphicx}
\usepackage{color}

\newcommand{\ha}{\hbox{H$\alpha$}}
\newcommand{\hb}{\hbox{H$\beta$}}
\newcommand{\oii}{\hbox{[O\,{\sc ii}]}}
\newcommand{\oiii}{\hbox{[O\,{\sc iii}]}}

\newcommand{\nii}{\hbox{[N\,{\sc ii}]}}

\newcommand{\mnras}{MNRAS}
\newcommand{\apj}{ApJ}

\newcommand{\aap}{A\&A}
\newcommand{\aj}{AJ}
\newcommand{\apjl}{ApJL}

\title[Environmental dependence of metallicity gradient]{SDSS-IV MaNGA: Environmental dependence of gas metallicity gradients in local star-forming galaxies}
	  
\author[J. Lian et al.]
{Jianhui Lian$^1$\thanks{jianhui.lian@port.ac.uk},
	Daniel Thomas$^1$, Cheng Li$^2$, Zheng Zheng$^3$, Claudia Maraston$^1$,
	\newauthor 
	Dmitry Bizyaev$^{5,6}$, Richard R. Lane$^{7}$, Renbin Yan$^{8}$\\
	$^1$Institute of Cosmology and Gravitation, University of Portsmouth, Burnaby Road, Portsmouth, PO1 3FX, UK\\
	$^2$Department of Astronomy, Tsinghua University, Beijing 100084, China\\
	$^3$National Astronomical Observatories of China, Chinese Academy of Sciences, 20A Datun Road, Chaoyang District, Beijing 100012, China\\
	$^4$Apache Point Observatory, P.O. Box 59, Sunspot, NM 88349\\
	$^5$Apache Point Observatory and New Mexico State
	University, P.O. Box 59, Sunspot, NM, 88349-0059, USA\\
	$^6$Sternberg Astronomical Institute, Moscow State
	University, Moscow, Russia\\
	$^7$Pontificia Universidad Cat\'olica de Chile,
	Instituto de Astrofısica,
	Av. Vicuna Mackenna 4860, 782-0436 Macul, Santiago, Chile\\
	$^{8}$Department of Physics and Astronomy, University of Kentucky, 505 Rose Street, Lexington, KY 40506, USA
}

\begin{document}
	
\maketitle

\begin{abstract}
Within the standard model of hierarchical galaxy formation in a $\Lambda$CDM Universe, the environment of galaxies is expected to play a key role in driving galaxy formation and evolution. In this paper we investigate whether and how the gas metallicity and the star formation surface density ($\Sigma_{\rm SFR}$) depend on galaxy environment. To this end we analyse a sample of {1162} local, star-forming galaxies from the galaxy survey Mapping Nearby Galaxies at APO (MaNGA). 
Generally, both parameters do not show any significant dependence on environment. However, in agreement with previous studies, we find that low-mass satellite galaxies are an exception to this rule. The gas metallicity in these objects increases while their $\Sigma_{\rm SFR}$ decreases slightly with environmental density. The present analysis of MaNGA data allows us to extend this to spatially resolved properties. Our study reveals that the gas metallicity gradients of low-mass satellites flatten and their $\Sigma_{\rm SFR}$ gradients steepen with increasing environmental density. By extensively exploring a chemical evolution model, we identify two scenarios that are able to explain this pattern: metal-enriched gas accretion or pristine gas inflow with varying accretion timescales. The latter scenario better matches the observed $\Sigma_{\rm SFR}$ gradients, and is therefore our preferred solution. In this model, a shorter gas accretion timescale at larger radii is required. This suggests that `outside-in quenching' governs the star formation processes of low-mass satellite galaxies in dense environments. 
\end{abstract}	


\section{Introduction}
In modern galaxy formation models galaxy environment is expected to play a critical role in the evolution of galaxies. In the $\Lambda$CDM model \citep{white1978,davis1985}, mergers of galaxies and their dark matter haloes are among the key drivers of galaxy formation. Galaxies residing in denser environments are generally more subjected to merger events. Further key processes for galaxy evolution in dense environments such as galaxy clusters are tidal stripping \citep{read2006}, galaxy harassment \citep{farouki1981}, and strangulation of gas accretion from the intergalactic medium \citep{larson1980,peng2015}. In contrast, galaxies living in under-dense regions are not subjected to these processes and are therefore expected to follow a different evolutionary path compared to the galaxies in dense environment. As a consequence, much effort has been put into looking for observational signatures of an environmental dependence of galaxies properties. 

Metal enrichment is one of the fundamental evolutionary processes in galaxies. While the gas phase metallicity characterises the most recent evolutionary path, stellar metallicity carries the record of the entire formation history. A strong correlation between gas metallicity and stellar mass is well established in the local \citep{lequeux1979,tremonti2004} and high-redshift Universe \citep{savaglio2005,erb2006, maiolino2008}. 
A similar but steeper scaling relation between the stellar metallicity and stellar mass has also been found in the local Universe \citep{thomas2010,peng2015,lian2018a}.

The gas metallicity is found to also depend on other galaxy properties such as star formation rate (SFR, \citealt{ellison2008,laralopez2010,mannucci2010}), age \citep{lian2015}, or gas mass fraction \citep{hughes2013}. While these observational results suggest that gas metallicity is largely driven by internal processes, metal enrichment will also be affected by environmental processes such as gas inflow and outflow \citep{tinsley1980,lilly2013}. Studying the environmental dependence of gas metallicity therefore sheds light on how galaxy evolution is affected by the environment and provides valuable constraints on the galaxy formation theory. 

Plenty of studies exist in the literature investigating the environmental dependence of integrated gas and stellar metallicity in local galaxies. Generally the effect of environment is found to be weak or not present. 
A slightly higher gas metallicity has been found in galaxies that reside in denser environment \citep{mouhcine2007,cooper2008,ellison2009,petropoulou2012,peng2014,wu2017}. The difference in gas metallicity of galaxies in dense and under-dense regions are generally small ($<0.1$ dex). \citet{peng2014} further separate between central and satellite galaxies to investigate the environmental dependence of gas metallicity. They find that higher gas metallicities are found in galaxies residing in denser environments, but only for satellite galaxies. The effect remains subtle, though, and several other studies find no significant difference in gas metallicity between galaxies in different environments \citep{kacprzak2015,pilyugin2017}. 

So far, relatively little attention has been paid to the metallicity distribution within galaxies and its environmental dependence. Understanding whether and how the galaxy environment affects radial gradients in metallicity will provide further insight on the interplay between galaxies and their environment. Integral field unit (IFU) spectroscopy provides an ideal observational strategy to derive the spatial distribution of metallicity in galaxies. Based on many recent and on-going IFU surveys, including CALIFA \citep{sanchez2012}, SAMI \citep{croom2012} and MaNGA \citep{bundy2015}, a number of recent studies have been carried out on the radial gradient of both gas and stellar metallicity in local galaxies (\citealt{sanchez2014,ho2015,perez2016,sanchez2016,belfiore2017,goddard2017,lian2018b}, Schaefer et al, in preparation).

Among them, Schaefer et al (in preparation) study the dependence of gas metallicity of satellite galaxies on the local environment and found a higher gas metallicity of satellites residing in a halo with a more massive central galaxy.

The goal of this work is to perform a systematical analysis of the potential connection between the radial gradient of gas metallicity with galaxy environment. Our study is based on the MaNGA IFU survey which provides currently the largest IFU galaxy sample with exquisite mapping of environmental density. The present study is being carried out in parallel with a study by another research group in MaNGA on the dependence of gas metallicity on the local environment and halo mass (Schaefer et al, in preparation). The observational results of both studies are in good agreement, and we will provide a more detailed comparison in the Results section. 

The structure of the paper is as follows. We will introduce the sample selection and determination of gas metallicity, SFR, and galaxy environment in \textsection2. A qualitative analysis followed by a quantitative assessment will be provided in \textsection3. In \textsection4 we will discuss the possible scenarios proposed to explain the observations by means of a chemical evolution model. The conclusions are presented in \textsection5. Throughout this paper, we adopt the cosmological parameters, $H_0$ = 71${\rm kms^{-1}Mpc^{-1}}, \Omega_{\Lambda}=0.73$ and $\Omega_{\rm m}= 0.27$.   

\section{Data}
MaNGA, one of the three core projects of the SDSS fourth generation survey \citep{blanton2017}, aims at mapping 10,000 local galaxies at completion in 2020 \citep{bundy2015} with spatially-resolved spectroscopy observations. The spectra cover a wide range in wavelength from 3,600\AA\ to 10,300 \AA\ at a resolution of $R\sim2000$. For galaxies with different sky coverage, 5 different fibre bundles with different number of fibers are used to ensure each galaxy is covered out to at least 1.5 $r_{\rm e}$ \citep{law2015,yan2016a,wake2017}. The raw spectra are reduced by the Data Reduction Pipeline (DRP, \citealt{law2016}) for wavelength calibration, flat-field correction, sky subtraction, and relative flux calibration at a $\sim1\%$ photometric uncertainty \citep{yan2016b}. Then the reduced data cubes are analysed by the Data Analysis Pipeline (DAP, \citealt{westfall2019}), the main survey-level software package to obtain further information such as kinematics and emission/absorption-line fluxes. To achieve a signal-to-noise (SNR) ratio of the spectral continuum above 10, spaxels with ${\rm SNR}<10$ are binned based on a Voronoi binning algorithm \citep{cappellari2003}. 

\subsection{Sample selection}
The galaxy sample of this work is selected from the seventh of MaNGA Product Launches (MPL-7), which contains 4652 galaxies and is part of SDSS-IV Data Release 15 \citep{aguado2018}. There are two galaxy samples observed by MaNGA following a 2:1 split. The `Primary' sample contains galaxies with IFU observations out to $1.5 r_{\rm e}$, while the 'Secondary' sample provides a wider coverage out to 2.5 $r_{\rm e}$. Both these two galaxy samples are used for the sample selection in this work.  
To obtain a robust measurement of the metallicity radial gradient, we require each galaxy to have at least four valid measurements of gas-phase metallicity, two within $0.8 r_{\rm e}$ and two beyond $0.8 r_{\rm e}$. Valid measurement means the spectrum has an SNR in the emission lines \oii$\lambda3727$, \hb, \oiii doublet, \ha, and \nii$\lambda6584$ above 5 and emission line ratios classified as star-forming to the demarcation line proposed by \citet{kewley2001} in the BPT diagram \citep{baldwin1981}. 

{There is now growing evidence that diffuse ionized gas (DIG) can affect the metallicity gradient measurement  \citep{zhang2017,poetrodjojo2019}. DIG is generally found in low \ha\ surface brightness regions which show different emission line ratios from normal HII regions. However, the selection of star forming regions through the BPT diagram as discussed above helps to largely exclude DIG regions. To further exclude spaxels that are potentially affected by DIG, we apply an additional criterion by only selecting spaxels with [SII]/\ha$>0.29$ as suggested by \citet{poetrodjojo2019}.} With these selection criteria {$1,162$} galaxies are selected.

\begin{figure}
	\centering
	\includegraphics[width=9cm]{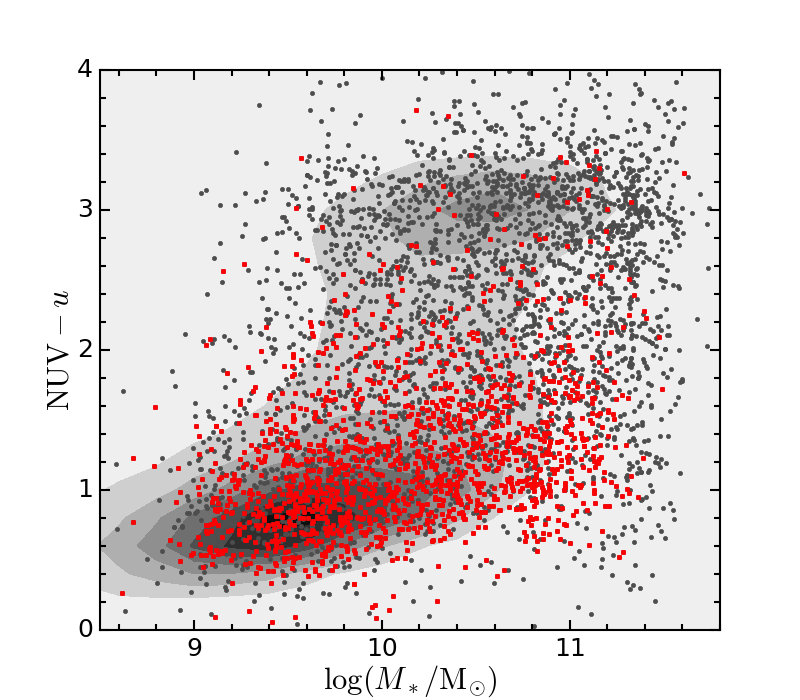}
	\caption{Galaxy distribution in the ${\rm NUV}-u$ colour-mass diagram. Grey dots are galaxies observed by MaNGA while red squares are the galaxies selected in this work. The underlying contour indicates the distribution of SDSS galaxies.  
	}
	\label{mass-nuvu}
\end{figure}

The stellar masses of these galaxies are taken from the NASA-Sloan-Altas (NSA) catalog \citep{blanton2011} while the emission line fluxes are from the MaNGA DAP where each emission line is fitted with a Gaussian profile. 
To account for the difference in Hubble constant which was assumed to be 100 ${\rm kms^{-1}Mpc^{-1}}$ when NSA mass was calculated, we multiple the mass by a factor of 2. 
A Voronoi binning scheme is used to bin the spectra of individual spaxels to achieve a signal-to-noise ratio in the stellar continuum higher than 10 \citep{cappellari2003}. Elliptical polar coordinates of each Voronoi cell are also adopted from the MaNGA DAP. Galactic reddening has been considered assuming the reddening law from \citet{donnell1994} {with $R_{\rm V}=3.1$}. To correct for internal dust extinction, we adopt the Balmer decrement method assuming case B recombination with an initial $\ha/\hb=2.86$ and an extinction law from \citet{cardelli1989}. 
Figure~\ref{mass-nuvu} shows the distribution of galaxies in the ${\rm NUV}-u$ colour-mass diagram. The grey dots are all MaNGA galaxies while the red squares are the galaxies in our sample. The underlying contours indicate the distribution of SDSS galaxies. Most galaxies selected in this work are located in the so-called `blue cloud' region, suggesting they are actively forming stars.    
 
\subsection{Gas metallicity and SFR}
Many methods have been proposed to measure the gas-phase metallicity of emission-line galaxies or HII regions \citep{kewley2008,maiolino2008,curti2017}. One of the most reliable methods is called `T$_{\rm e}$' method for which the electron temperature in the ionized gas is required. The relative strength of two excitation lines of the same ion is a good indicator of electron temperature. For example, \oiii$\lambda4363$ and \oiii$\lambda5007$ is one of the widely-used emission line pairs. However, the auroral line \oiii$\lambda4363$ is usually very weak, especially in metal-rich galaxies. Therefore many alternative empirical methods that calibrate emission line ratios to the metallicity derived by the `T$_{\rm e}$' method have been proposed, such as N2=\nii/\ha, $R_{23}$=(\oii$\lambda3727$+\oiii$\lambda4959$+\oiii$\lambda5007$)/\hb, O3N2=(\oiii/\hb)/(\nii/\ha), and N2O2=\nii/\oii\ (e.g. \citealt{dopita2000,kewley2001,pettini2004,pilyugin2005}). 

Although these empirical calibrations are useful in determining the gas metallicity, they face many challenges. {The scatter in metallicity for each calibrator is generally small. However,} different calibrations usually derive values that differ by up to 0.7 dex \citep{shi2005,kewley2008}. The physical origin of this discrepancy is still not fully understood. Moreover, these calibrations which are based on integrated observations of HII regions or emission-line galaxies face problems when applied to spatially-resolved observations. Using MaNGA IFU observations, \citet{zhang2017} found that regions with low \ha\ surface brightness in star forming galaxies, which are referred to as regions with diffuse ionized gas (DIG), show distinct emission line ratios compared to normal star-forming/HII regions.  
The distinctive emission line ratios suggest that the DIG, not like the HII regions ionized by massive stars, is ionized by evolved stars. 
By analyzing the impact of DIG on metallicity measurements based on emission line calibrations in detail, \citet{zhang2017} find that, among the various calibrations, N2O2=\nii/\oii\ is still a robust metallicity indicator even in the present of DIG. Moreover, \citet{hwang2019} show that the N2O2-based metallicity is not sensitive to the ionization parameter which affects many other metallicity calibrators. In light of these works, to minimize the impact from DIG in the metallicity measurements, we adopt the N2O2 calibration $12+{\rm log(O/H)=8.93+log}(1.54020+1.26602\times R+0.167977\times R^2),\ R={\rm log(\nii/\oii)}$ from \citet{dopita2013} which was also used in \citet{zhang2017} and \citet{hwang2019} to derive gas metallicity. 

{As discussed in \textsection2.2, spaxels potentially affected by DIG regions are excluded to a large extent based on cuts using multiple emission line ratios. Our selected spaxels are therefore dominated by star formation regions, so that we can estimate the SFR based on emission line luminosity.} For each spaxel, we derive the SFR based on the extinction-corrected \ha\ luminosity using the SFR calibration of \citet{kennicutt1998}.    


\subsection{Environment}
The environment of galaxies is usually considered to be the density field in which they reside. Many approaches have been proposed to quantify galaxy environment, such as number density of a certain volume around the galaxy in question or the strength of gravitational interaction (i.e. tidal strength) imposed by nearby galaxies. A comprehensive compilation of quantification of the environment for MaNGA galaxies is conducted by Argudo-Fern{\'a}ndez, M. which includes galaxy group classification from \citet{yang2007}, local densities and tidal strength within various volume size calculated by different teams \citep{argudo2015,etherington2015,wang2009,wang2012,goddard2017}, and galaxy pairs identified by the MaNGA-mergers working group.  

\begin{figure}
	\centering
	\includegraphics[width=\columnwidth]{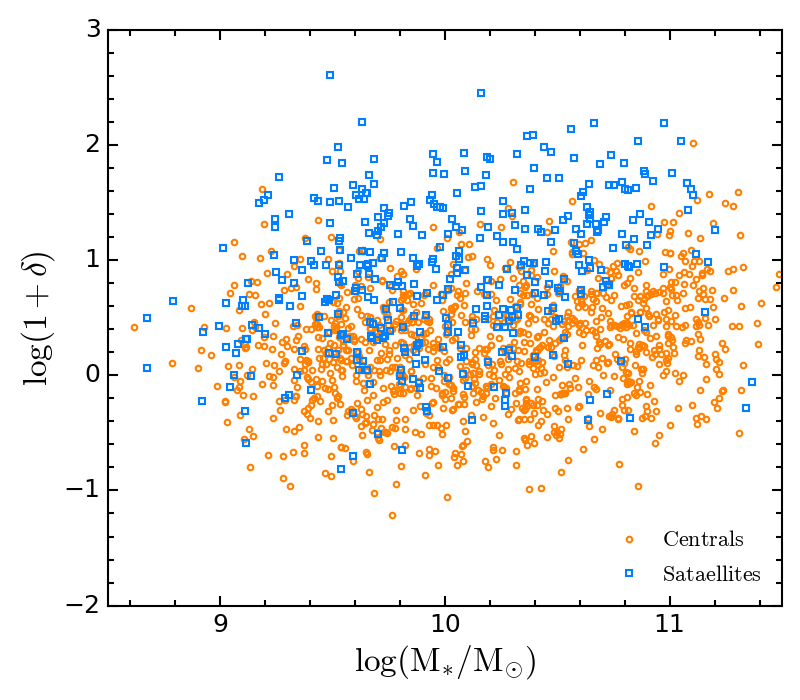}
	\caption{Distribution of the central (orange circles) and satellite galaxies (blue squares) in our galaxy sample in the mass-overdensity diagram.  
	}
	\label{mass-density}
\end{figure}

In this work, we adopt two approaches to quantify galaxy environment. We first use the galaxy group classification from \citet{yang2007} to divide our galaxy sample into central galaxies and satellite galaxies (centrals and satellites hereafter). To further quantify the local environment, we adopt the $N^{\rm th}$ Nearest neighbour method \citep{muldrew2012}. The density within the $4^{\rm th}$ or $5^{\rm th}$ nearest neighbour is considered to be representative of the local environment as illustrated in \citet{baldry2006}. In this work we adopt the number density within the $5^{\rm th}$ nearest neighbour to represent the galaxy local environment which is calculated based on the algorithm developed in \citet{etherington2015}. 
The local overdensity is then defined as: 
\begin{equation} \label{eq1}
\delta=\frac{\rho-\rho_{\rm m}}{\rho_{\rm m}},
\end{equation}
where $\rho_{\rm m}$ is the mean density within a large volume centred on the target galaxy.  
Conventionally, a form of $\rm {log (1+\delta)}$ is usually used to express environmental density. Figure~\ref{mass-density} shows the distribution of the centrals (orange circles) and satellites (blue squares) in our galaxy sample in the mass-overdensity diagram. 
As expected, there is a weak correlation between the two quantities such that more massive galaxies tend to reside in denser regions. Compared to the centrals, satellites tend to  reside in denser environments. This is because a large fraction of central galaxies are actually isolated systems. At the high-mass end ($>10^{10.5}M_{\odot}$), most galaxies tend to be centrals. This is a natural result of the definition of the central galaxy which is the most massive galaxy in a given galaxy group.   

As an additional measure for environment, we also use the galaxy environment measured in the reconstructed density field from $N$-body simulations performed by \citet{wang2009,wang2012}. In particular, we use the overdensity measured within 3Mpc 
from each target galaxy to indicate the local galaxy environment. The results based on this density measurement are shown in the Appendix. This test confirms that the conclusions of this paper are not affected by the exact choice of method for the measurement of environmental density. 

\section{Results}
\begin{figure*}
	\centering
	\includegraphics[width=16cm]{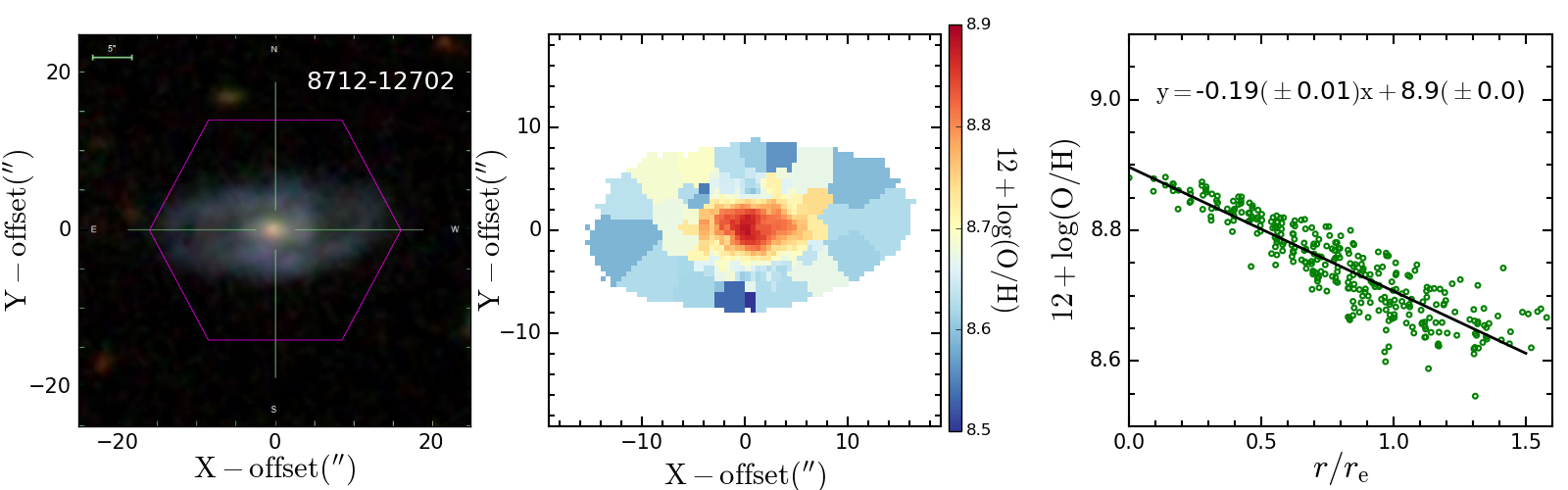}
	\caption{Left panel: Optical SDSS image with the galaxy ID shown in the top-right corner. Middle panel: Two dimensional spatial distribution of gas metallicity for this galaxy. Right panel: One dimensional radial distribution of gas metallicity. A linear relation is fit to the radial distribution (black line). The equation of the best-fit line is shown at the top.
	}
	\label{example}
\end{figure*}

In this section we present the results from the MaNGA data analysis. 
{Figure~\ref{example} shows the metallicity distribution of a normal MaNGA galaxy as an example. The left-hand panel shows the optical SDSS image of this galaxy. The ID of this object is shown in the top-right corner. The 2-dimensional spatial distribution of gas metallicity of this galaxy is shown in the middle panel, while the 1-dimensional radial distribution is shown in the right-hand panel. A linear correlation is fit to the radial distribution as shown by the black line. The equation of the best-fit line is shown at the top.}

\subsection{Metallicity radial profile}
It is well known that the gas metallicity of galaxies is strongly correlated with the galaxy mass (i.e. the mass-metallicity relation, e.g. \citealt{tremonti2004}). 
To investigate the intrinsic dependence of gas metallicity on galaxy environment, it is important to separate the mass-metallicity and mass-environment relations. 

To this end, we split our central and satellite samples into three mass bins with stellar mass ${\rm log(M_*/M_{\odot})<9.7}$, $9.7<{\rm log(M_*/M_{\odot})<10.2}$, ${\rm log(M_*/M_{\odot})>10.2}$. To qualitatively assess the environmental effect on metallicity, we further split the centrals and satellites in each mass bin into quartiles with different environmental densities. $\delta<25$th percentile defines the low $\delta$ bin while $\delta>75$th percentile defines the high $\delta$ bin. The mid-low $\delta$ bin indicates the $25{\rm th}<\delta<50$th percentile while the mid-high $\delta$ bin represents the $50{\rm th}<\delta<75$th percentile. To obtain the radial profile of metallicity, for each galaxy, we separate the spaxels into eight radial bins with bin width of 0.2 re and obtain the median metallicity in each radial bin. A median radial profile is then obtained for galaxies in each mass and density bin. 

\begin{figure*}
	\centering
	\includegraphics[width=16cm]{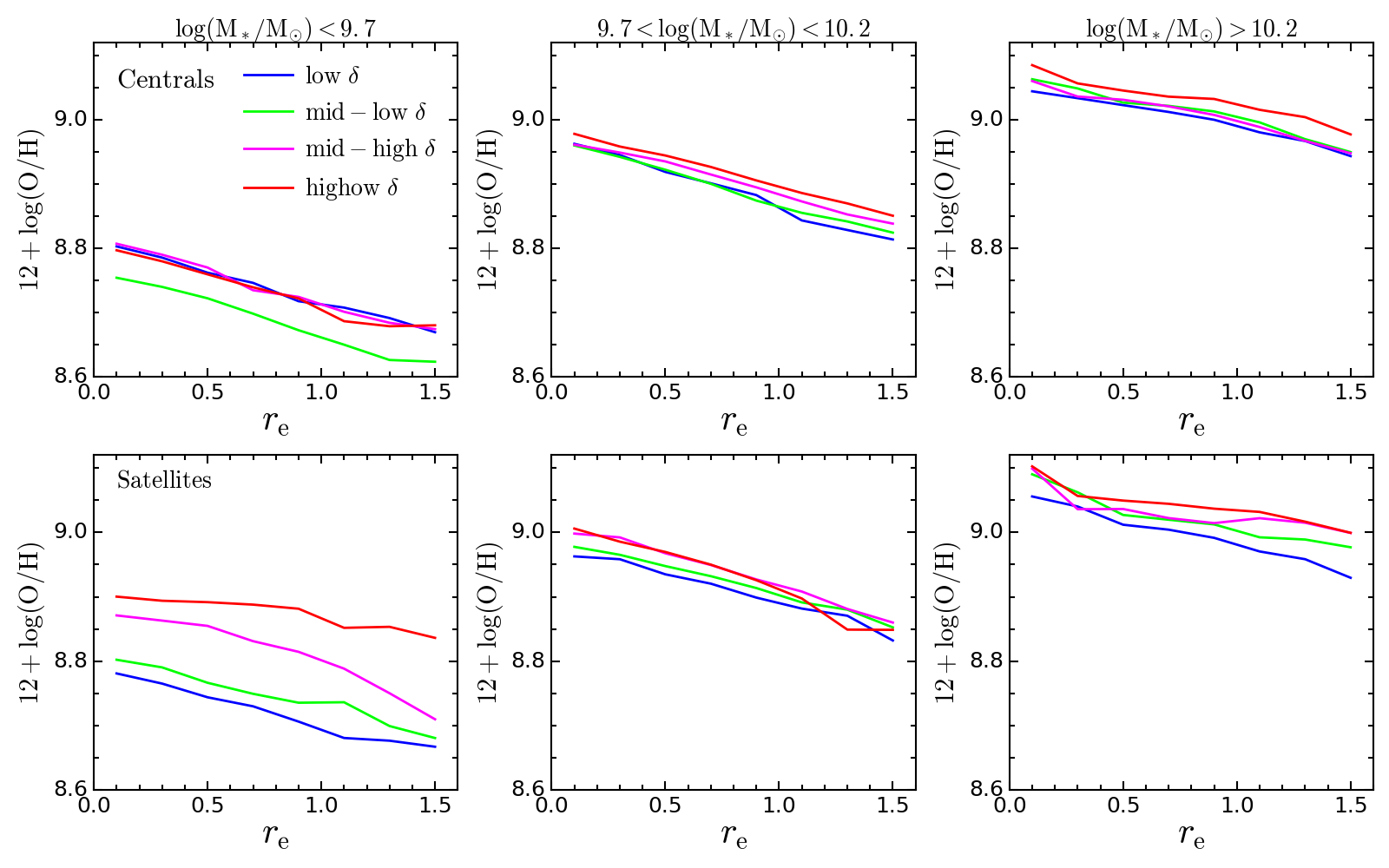}
	\caption{Radial profile of gas metallicity of central (top row) and satellite galaxies (bottom row) in each mass and environmental density bin. Environmental densities are obtained using the 5$^{th}$ Nearest neighbour method \citep{etherington2015}. Panels from left to right indicate the three mass bins from low to high mass. Galaxies residing in different environments are marked as different colours as shown in the legend in the top-left panel.  
	}
	\label{profile-obs}
\end{figure*}

Figure~\ref{profile-obs} shows the median radial profile of gas metallicity for centrals (top row) and satellites (bottom row) in each mass and density bin. 
Panels from left to right indicate the three mass bins (${\rm log(M_*/M_{\odot})<9.7}$, $9.7<{\rm log(M_*/M_{\odot})<10.2}$, ${\rm log(M_*/M_{\odot})>10.2}$) from low to high mass. Galaxies residing in different environmental densities are marked through different colours as illustrated in the legend in the top-left panel. 

It can be seen that gas metallicity in shows a universal negative radial gradient for all galaxies, regardless of the environment and stellar mass. This negative gradient in gas metallicity is in good agreement with previous literature (e.g. \citealt{sanchez2014}; \citealt{belfiore2017}). 
With galaxy mass increasing from left to right, it can be also seen that there is a clear trend of a higher gas metallicity in more massive galaxies, i.e. the well-known mass-metallicity relation. 

On top of this there seems to be a weak dependence of the zero point of gas metallicity on the local overdensity in each mass bin with slightly higher gas metallicity in denser environments. This trend may not be intrinsic but a result of residual dependence on stellar mass and the mass-environment correlation (see Figure~\ref{mass-density}). The much stronger dependence of gas metallicity on mass than environment suggests that the mass of a galaxy plays a much more important role than the environment in regulating the metal enrichment process in galaxies. This is consistent with the results found for the stellar populations \citep{thomas2010,goddard2017}.

The only exception appear to be low-mass satellites (bottom-left panel) that show a very clear trend with higher gas metallicities in denser environments. The difference in gas metallicity is of the order of 0.1 dex. This suggests that the dependence of gas metallicity on environment in low-mass satellites is indeed intrinsic rather than caused by a residual dependence on stellar mass. We note that Schaefer et al (in preparation) identify the same residual effect for low-mass satellites. This finding is in good agreement with the trend of higher global gas metallicity of satellites in denser environment and the lack of any environmental dependence for centrals found by \citet{peng2014} based on SDSS integrated fibre spectroscopy observations. 

The spatially resolved data of MaNGA allows us to take this result one step further. In addition to the dependence of global gas metallicity on environment in the low-mass satellites, it is interesting to note there is a hint of flatter gas metallicity gradients in denser environment in Figure~\ref{profile-obs}. This interesting trend is found for the first time and will be quantified in \textsection3.3.

To examine whether our results depend on the methodology of environment determination, we show the radial profile of gas metallicity with the overdensity calculated based on the density field reconstructed by $N-$body simulation in Figure~\ref{profile-lss} in the Appendix. Consistent trends are found. 

\subsection{$\Sigma_{\rm SFR}$ radial profile}

\begin{figure*}
	\centering
	\includegraphics[width=16cm]{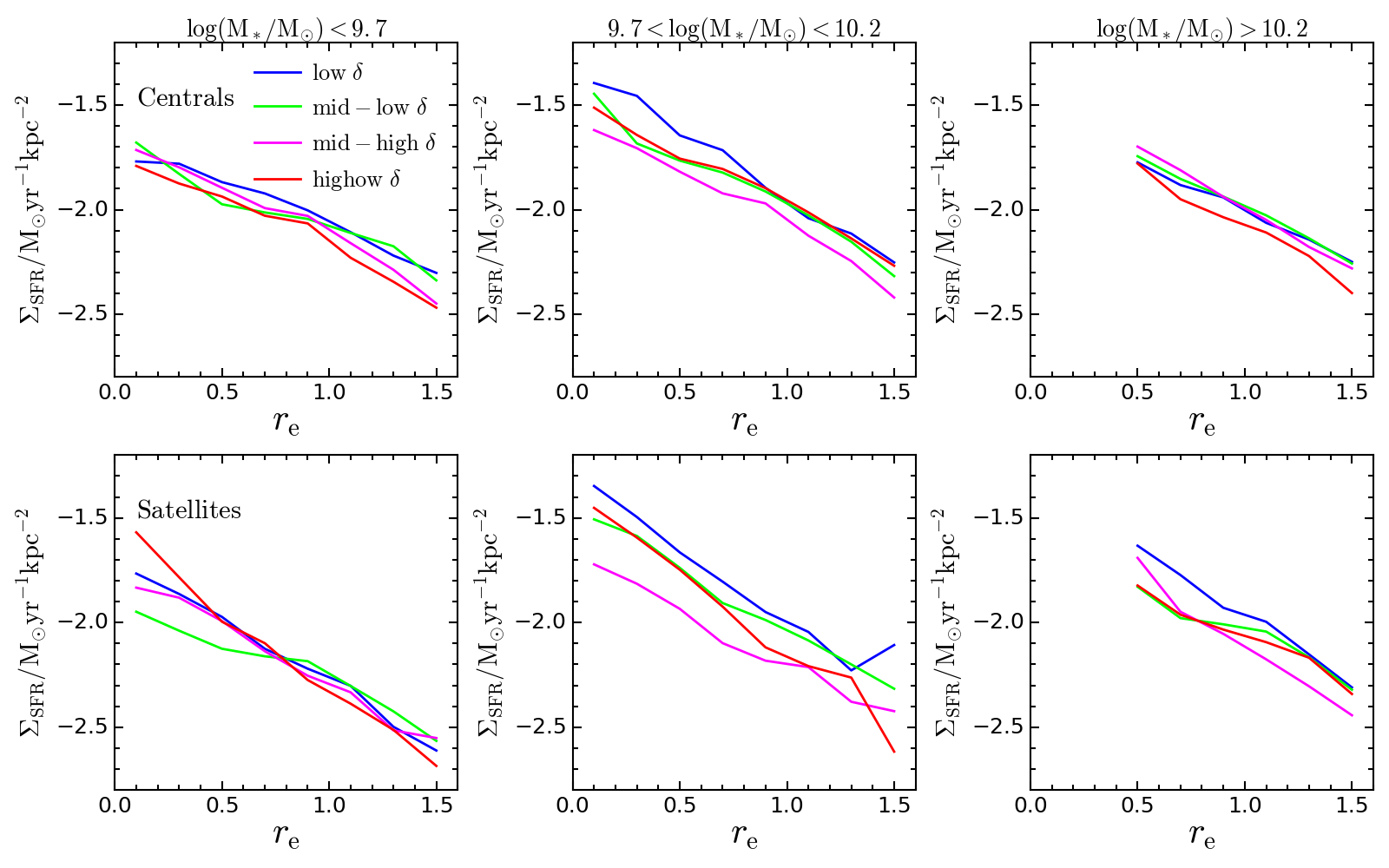}
	\caption{Similar to Figure~\ref{profile-obs} but showing SFR surface density instead of gas metallicity.   
	}
	\label{sfr-profile-obs}
\end{figure*}

Similar to the gas metallicity, we derive the radial profile of SFR surface density $\Sigma_{\rm SFR}$ as a function of galaxy stellar mass and environmental density (Figure~\ref{sfr-profile-obs}). 
There is a slight dependence of SFR on environment with slightly lower SFR in denser environments. This weak trend may be caused a residual dependence of specific SFR on stellar mass (i.e. the mass-SFR relation, e.g. \citealt{noeske2007}).  
Different from the case of gas metallicity, low-mass satellites do not deviate from this general trend. There only is a hint for a trend of a steeper $\Sigma_{\rm SFR}$ gradient in denser environments. Interestingly, a similar trend is found in \citet{schaefer2017} based on the SAMI IFU survey. More recently, \citet{spindler2018} also found this trend based on an earlier release of MaNGA data but claimed that the trend may not be robust due to the poor statistic in the dense environment. 

\subsection{Quantitative analysis} 
In addition to the qualitative assessment presented in the previous sections, we now turn to conduct a more quantitative analysis. Instead of splitting galaxies into four discrete groups of environmental density, we now assess the environmental dependence directly considering the continuous distribution of galaxy properties as a function of environmental densities. 

\subsubsection{Gas metallicity}
{For each galaxy, as illustrated in the right panel of Fig~\ref{example}, we measure the metallicity gradient by linearly fitting the metallicity measurements as a function of radius. We then use the metallicity from this fit at $0.8\; r_{\rm e}$ to calculate the global metallicity ${\rm Z_{0.8}}$ (normalized to solar abundance).}



\begin{figure*}
	\centering
	\includegraphics[width=18cm]{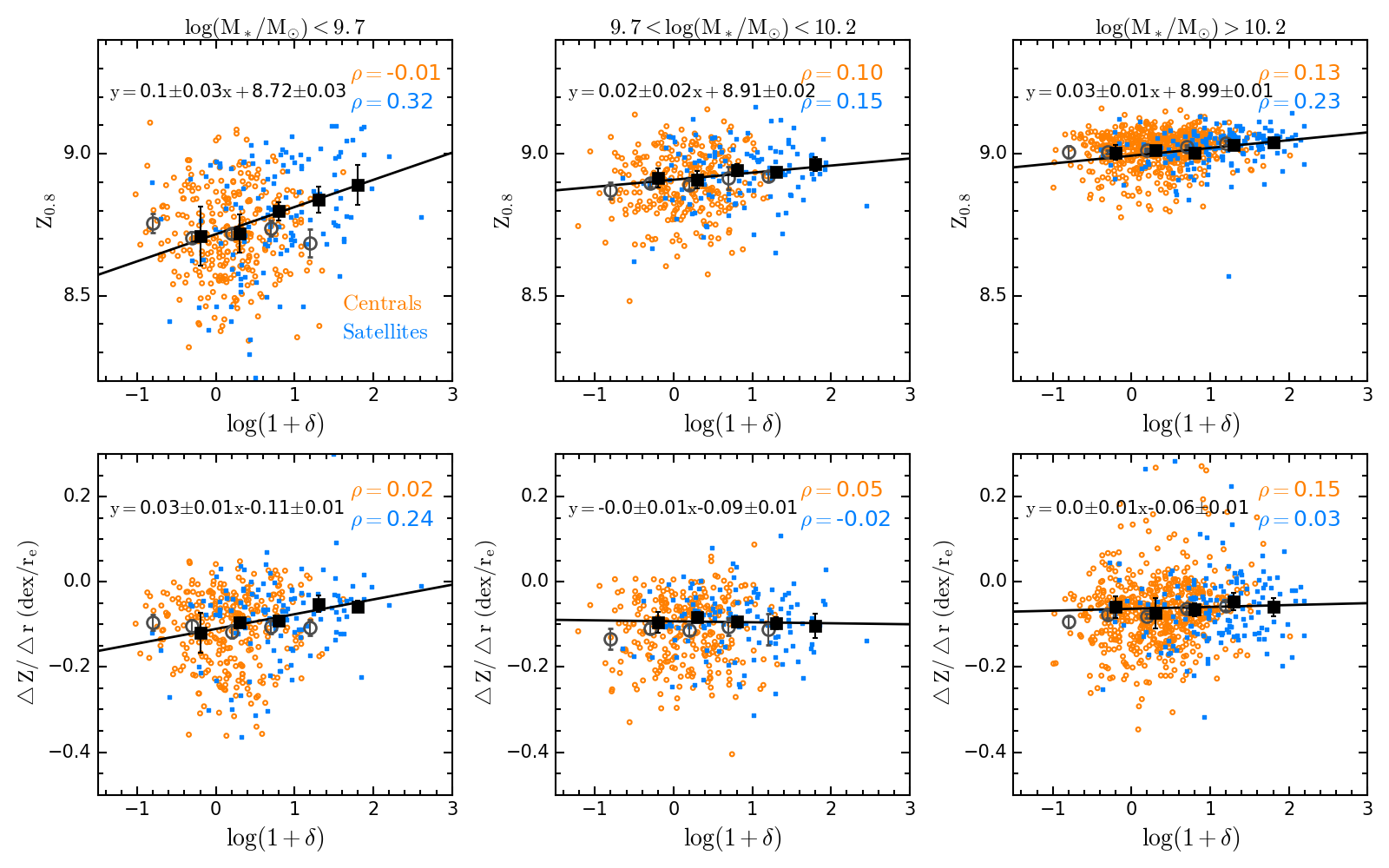}
	\caption{Global metallicity (top row) and metallicity gradient (bottom row) of centrals (orange circles) and satellites (blue squares) as a function of environmental density and galaxy mass. The environmental densities are measured through the 5$^{\rm th}$ nearest neighbour method. Panels from left to right indicate three bins of increasing mass. Black and grey error bars indicate the median and the error of the median global metallicity and metallicity gradient in five equally spaced environmental density bins for satellites and centrals, respectively. The quantities shown in the top-right corner in each panel indicate the Pearson's correlation coefficient.  
	}
	\label{absm-obs}
\end{figure*}

Figure~\ref{absm-obs} shows the global metallicity (top row) and metallicity gradient (bottom row) of centrals (orange circles) and satellites (blue squares) as a function of environmental density. Similar to our previous qualitative analysis, galaxies are split into three mass bins to minimise an additional dependence introduced by galaxy mass. Each column indicates one mass bin from low mass on the left to high mass on the right.

To illustrate the overall trend, we calculate the median and error of median measurements in five equally spaced environmental density bins with bin width of 0.5. Considering the slightly denser environment the satellites reside compared to the centrals, the density range for binning is shifted by 0.5 to higher densities for satellites.    
Black and gray error bars indicate the results for satellites and centrals, respectively. To avoid overlap in the four common density bins, we slightly shifted the symbols on the x-axis between centrals and satellites. 

To quantify the strength of the correlation we calculate the Pearson correlation coefficient 
as shown in the top-right corner of each panel. It is evident that the correlation between global metallicity and environment is significant for low-mass satellites ($\rho=0.31$), but much weaker or non-existent for more massive satellites and centrals of all masses ($0.05\leq\rho\leq 0.21$). 

The black solid lines show linear best-fits for the satellites. As expected the slope of the correlation between global gas metallicity and environmental overdensity is strongest and most significant for low-mass satellites with a slope of $0.09\pm0.02$. Hence the correlation is significant at the 4$\sigma$ level, compared to only 1$\sigma$ for more massive satellites.      
The bottom panels of Figure~\ref{absm-obs} show the metallicity gradient. Again, there is a significant positive correlation only in low-mass satellites with a Pearson's coefficient of $\rho=0.24$, compared to $0.02\leq\rho\leq 0.13$ for all other cases. This implies that low-mass satellites residing in denser environments tend to have flatter gas metallicity gradients. The linear fit yields a slope of $0.04\pm 0.01$, implying a significance at the 4$\sigma$ level.

   
\subsubsection{SFR}
\begin{figure*}
	\centering
	\includegraphics[width=18cm]{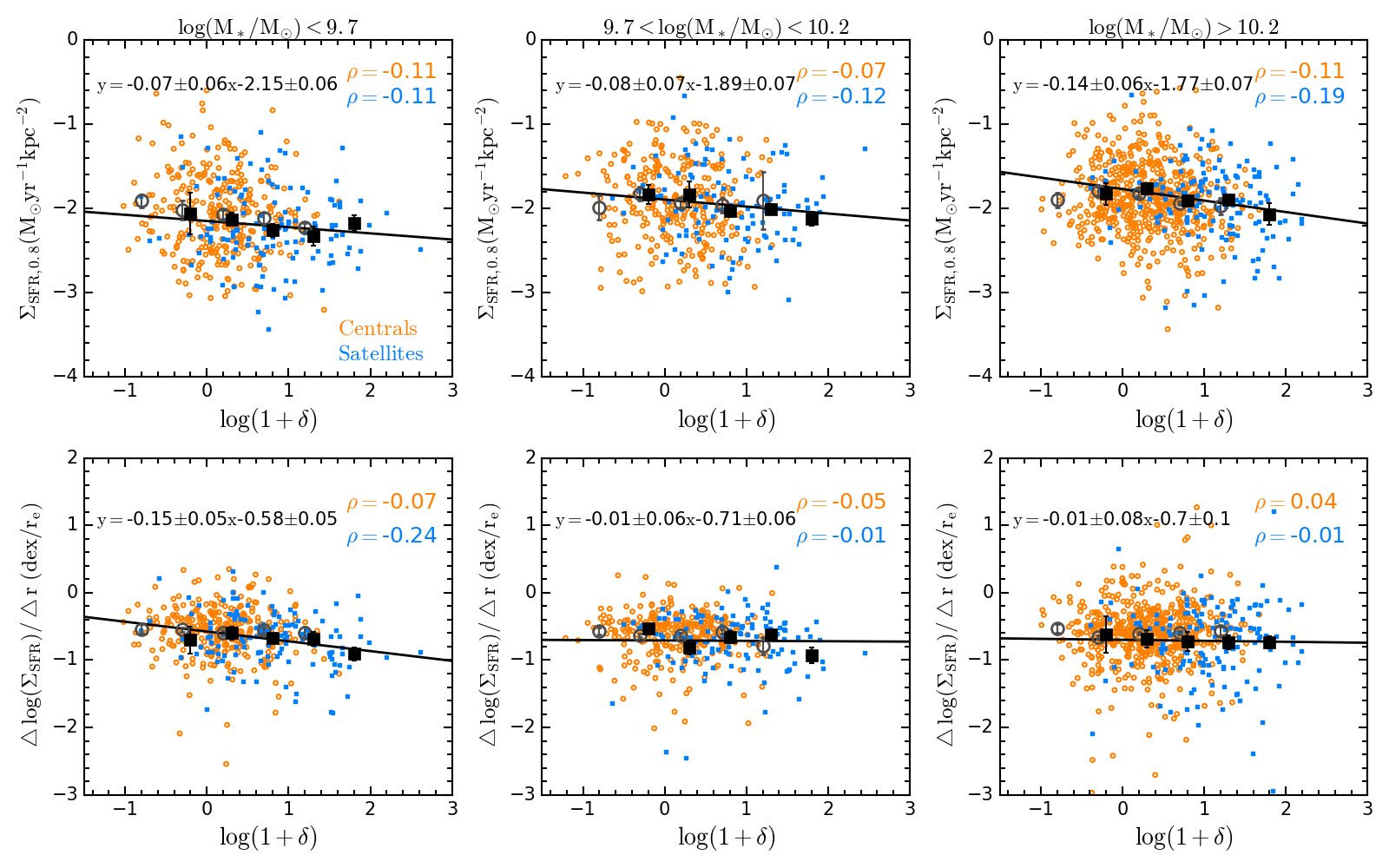}
	\caption{Similar to Figure~\ref{absm-obs} but showing the SFR surface density $\Sigma_{\rm SFR}$ instead of local gas metallicity. 
	}
	\label{sfr-absm-obs}
\end{figure*}

Following the same approach as for gas metallicity, we derive a global $\Sigma_{\rm SFR}$ at 0.8$r_{\rm e}$ and the $\Sigma_{\rm SFR}$ gradient. Figure~\ref{sfr-absm-obs} shows $\Sigma_{\rm SFR,0.8}$ (left-panel) and $\Sigma_{\rm SFR}$ gradient (right-panel) as a function of environmental densities. 
The correlation between $\Sigma_{\rm SFR}$ and $\Sigma_{\rm SFR}$ gradient with environment density are generally weak with relatively low Pearson's correlation coefficients ($0.1\leq\rho\leq 0.15$). The only exception is the $\Sigma_{\rm SFR}$ gradient in low mass satellites, for which we detect a weak correlation with $\rho=0.21$. The linear fit implies a significance at the 2$\sigma$ level.

Although the dependence of the global $\Sigma_{\rm SFR}$ on environmental density is not significant, similar trends of lower $\Sigma_{\rm SFR}$ and steeper negative $\Sigma_{\rm SFR}$ gradients in satellites residing in denser environments have been found in the literature \citep{schaefer2017,finn2018,spindler2018}.  

\section{Chemical evolution modelling}
To explore the physical origin of the correlations discussed in the previous section we employ a galactic chemical evolution model \citep{lian2018a}. It what follows we focus on {\it low-mass satellites}, for which the significant correlations of gas metallicity and $\Sigma_{\rm SFR}$ with environment have been found.

The model accounts for three basic processes: star formation, gas accretion and gas outflow. The formation of stars from the gas reservoir follows the Kennicutt-Schmidt star formation law \citep{kennicutt1998} with the coefficient lowered by a factor of 5 to account for the relatively low star formation efficiency in low mass galaxies \citep{lian2018a}. A Kroupa IMF \citep{kroupa2001} is adopted. Metals synthesized by stars will be released at the end of their lifetime. For more details about the model we refer the reader to \textsection3 in \citet{lian2018a}. Based on our model, we successfully reproduced the integrated and spatially-resolved gas and stellar metallicity in local galaxies \citep{lian2018a,lian2018b}, as well as the cosmic evolution of the mass-gas metallicity relation from $z\sim3.5$ to $z\sim0$ \citep{lian2018c}. 

\subsection{Initial conditions}
To model the process of how gas metallicity becomes dependent on environment in low mass satellites, we make a few assumptions. 

In general the model is assumed to evolve from the very early Universe to the present day. However, the gas metallicity of a galaxy is most sensitive to recent evolutionary processes \citep{lian2018a}, suggesting the environmental dependence of gas metallicity has been introduced over the past few Gyr. Therefore, in this work we calculate models evolving from a lookback time of a few Gyr to the present. 

We assume that today's satellites originally resided in the same environment (i.e.\ field) and therefore share the same evolutionary  path at early times. The typical dynamical friction timescale of accreted satellites to merge with the central galaxy of the order of a few Gyr depending on the orbital parameters and mass ratio between satellite and host halo \citep{taffoni2003,boylan2008,jiang2008,gan2010}. Here we adopt a timescale of 2 Gyr and assume therefore that the observed environmental dependence has been introduced during the past 2 Gyr. 
The exact value for this timescale does not significantly affect the main result of this work. A test with a model adopting a timescale of 5 Gyr will be discussed in the Discussion section.

The initial conditions of the model at a lookback time of 2 Gyr are based on today's observations of satellites residing in low-density regions.

\subsubsection{Star formation law}
We assume that the evolution of satellites residing in the environment with zero overdensity is not affected by the environment and therefore could be used to probe the initial condition before they become satellites.  
We further assume a constant SFH {for the low-mass galaxies}, which is supported by local and intermediate-redshift observations of the mass-SFR relation \citep{noeske2007} and the color-magnitude diagram \citep{weisz2014}. {Given a constant SFH, the gas content is also a constant and the initial radial distribution of gas mass could then be} derived from {the locally observed} $\Sigma_{\rm SFR}$ and the KS star formation law. 
{The locally observed SFR radial profile with zero overdensity is ${\rm log(\Sigma_{\rm SFR})}=-0.70\times r/r_{\rm e}-2.31$}. 

\subsubsection{Gas accretion}
In addition to the initial gas reservoir, external gas accretion is another important factor to keep fueling the star formation. We assume the gas accretion rate declines exponentially with time. A fiducial gas accretion timescale of 10 Gyr is adopted which is broadly consistent with the timescale obtained by fitting mass-SFR relations at different cosmic epochs \citep{noeske2007}. The initial accretion rate is set to be identical to the star formation rate at a lookback time of 2 Gyr to fulfill the assumption of a constant SFH. Therefore, following the SFR, the gas accretion rate is radially dependent. The metal abundance in the external gas accretion and the e-folding time of the exponential declining are free parameters.

In \citet{lian2018a} we found that gas outflows are important at the earlier evolutionary stages in the evolution of low-mass galaxies, but have been insignificant in recent times. \citet{heckman2002} suggests that gas outflows are common in starbursts in local universe with $\Sigma_{\rm SFR}$ above 0.1 M$_{\odot}$ yr$^{-1}$ kpc$^{-2}$. Note that most of centrals and satellites in this work have a SFR surface density below this threshold. Therefore no gas outflow is assumed in the present work. 

\subsubsection{Radial distributions}
The initial radial distribution of gas metallicity is again taken from observations. We assume there was no evolution in the gas metallicity radial gradient in the past 2 Gyr for galaxies where the environment effect was not significant. Considering that the low-mass satellites residing in the lowest density environment today experienced the least effect of environment on the metal enrichment, we adopt the metallicity gradient of these satellites to represent the gradient of low-mass satellites at a lookback time of 2 Gyr.

Based on our model in \citet{lian2018a}, the evolution of gas metallicity in low-mass galaxies with $M_*=10^{9.2}{\rm M_{\odot}}$ is $\sim 0.2$ dex over the past 2 Gyr. Therefore, we obtain the radial profile of gas metallicity in low mass satellites at a lookback time of 2 Gyr using the radial profile of low-mass satellites in {zero overdensity environment today with the zero-point lowered by 0.2 dex.} 
For simplicity, a linear form is adopted for the initial radial profile of gas metallicity, 
$12+{\rm log(O/H)}=-0.10\times r/r_{\rm e}+8.76$. {The slope and interception are taken from low-mass satellites with overdensity around zero.}

To account for the chemical evolution of satellites in different environments, we calculate four series of models. Each contains eight independent models, corresponding to the eight radial bins in each galaxy. No material exchange between adjacent radial bins is assumed. 

\subsection{Possible scenarios}
The metal enrichment in galaxies is the result of the balance between in-situ stellar metal production, dilution by pristine/metal-poor gas accretion, and removal through galactic outflow. The latter two processes are directly associated to the environment and therefore more likely responsible for the observed dependence of gas metallicity on environment. Since outflows tend to be weak in the recent Universe \citep{lian2018a,concas2017}, environmentally dependent gas accretion is the mostly promising solution. 

\subsubsection{Enriched gas accretion}

\begin{figure*}
	\centering
	\includegraphics[width=14cm]{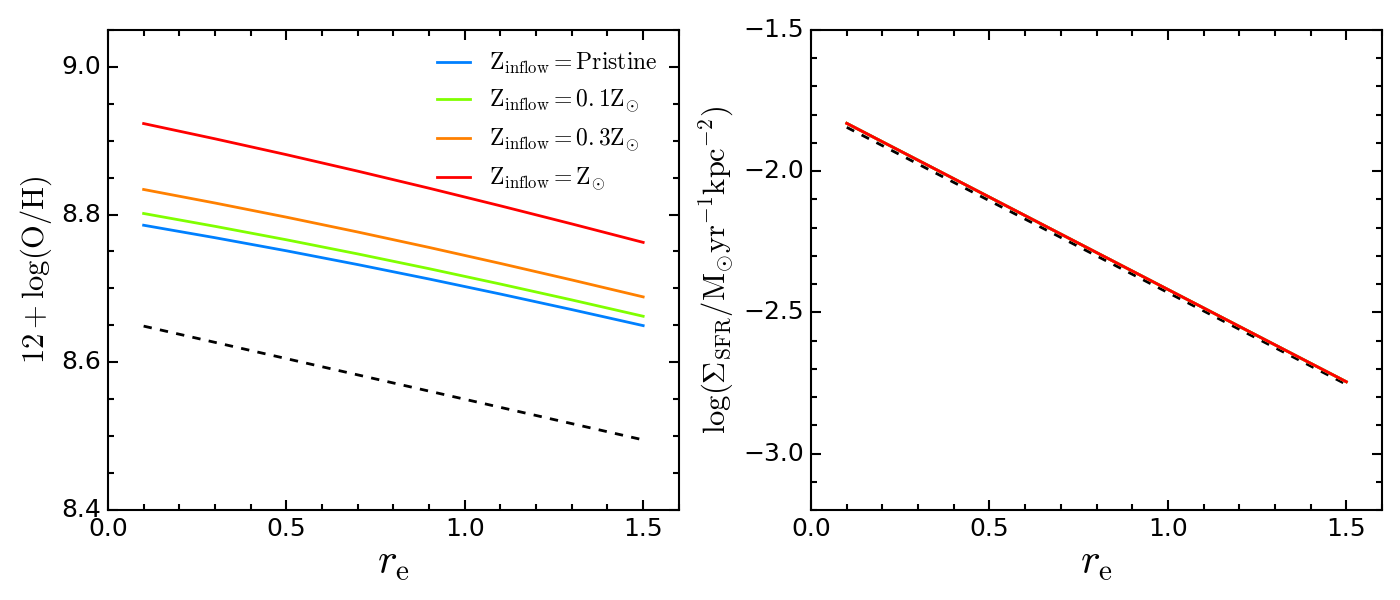}
	\caption{Prediction of the chemical evolution models with metal-enriched gas accretion. The dashed lines in both panels represent the initial condition at a lookback time of 2 Gyr. The solid lines with different colour indicate the present radial profile of metallicity (left panel) and $\Sigma_{\rm SFR}$ (right panel) predicted by the models with different metallicity in the accreted gas. The metallicity in the gas accretion is shown in the upper right corner in the left panel. In the right panel, solid lines with different colours overlap with each other as the star formation history in different models are the same. 
	}
	\label{gce-enrich}
\end{figure*} 

A scenario of enriched gas accretion was proposed by \citet{peng2014} to explain the dependence of global gas metallicity on environment in satellites. In this scenario, satellites residing in dense regions and hence being closer to the central galaxy in the same dark matter halo are more likely exposed to intergalactic medium that has been metal enriched by the central galaxy. By accreting this metal-enriched gas, the metal enrichment process in satellites will be accelerated, resulting in a higher gas metallicity of satellites in denser region. This scenario was proved to be successful to reproduce the environmental dependence of global gas metallicity in \citet{peng2014} based on their analytic chemical evolution model. Here we aim to test whether this scenario is able to also explain the environmental dependence of the metallicity gradient.

Figure ~\ref{gce-enrich} shows the prediction of the model with enriched gas accretion for the radial profile of gas metallicity (left panel) and $\Sigma_{\rm SFR}$ (right panel). Solid lines with different colours indicate the models with different pre-enrichment, from pristine to solar abundance. The four lines in the right panel completely overlap, i.e. the different model flavours predict identical radial profiles for $\Sigma_{\rm SFR}$. 
The dashed lines in the two panels indicate the initial radial profile of gas metallicity and $\Sigma_{\rm SFR}$ at a lookback time of 2 Gyr. 

The model with enriched gas accretion leads to higher gas metallicities than the one with pristine gas accretion. This ought to be expected, and is qualitatively consistent with the model prediction in \citet{peng2014}. More interestingly, the gas metallicity gradient does not show any significant variation.
Star formation rates, instead, do not vary in this scenario.


Note that the metallicity of the accreted gas is assumed to be radially independent. To match the observed trend of flatter gas metallicity gradients in denser environments (see Section 3), a radially dependent metal enrichment in gas accretion is required. The gas accreted at large radii would need to be more metal-rich in order to generate a flattening of the gradient.
 

\subsubsection{Variable gas accretion timescale}

\begin{figure*}
	\centering
	\includegraphics[width=14cm]{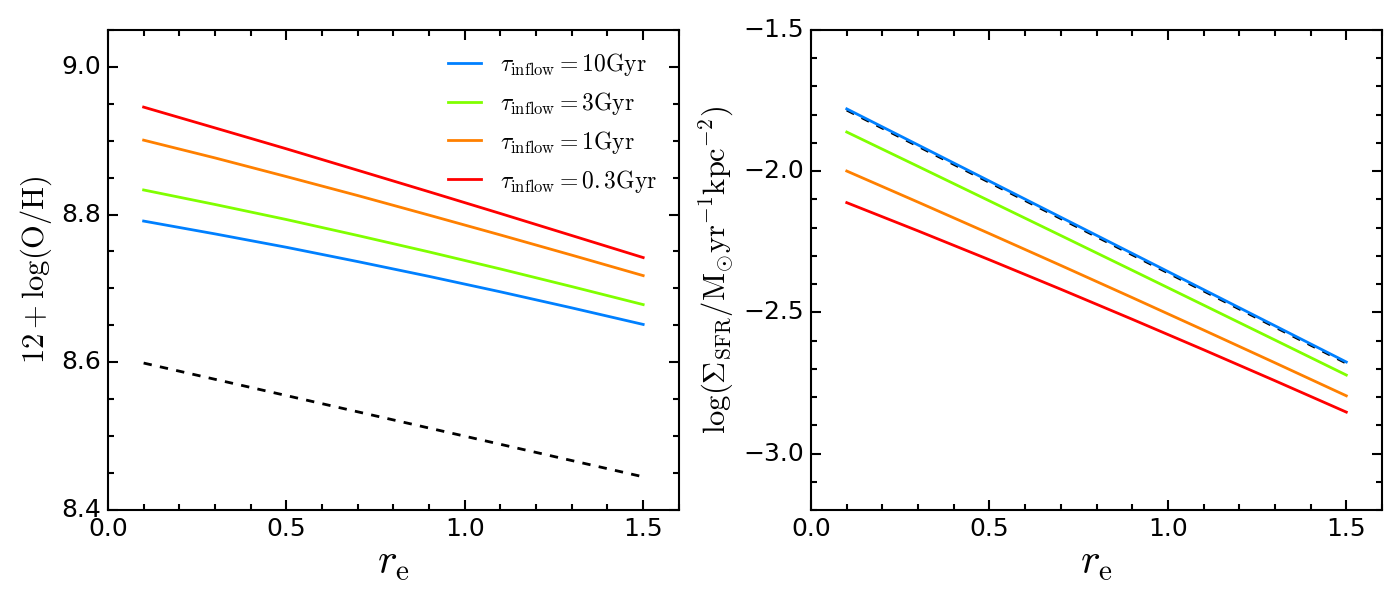}
	\caption{Similar to Figure~\ref{gce-enrich} but showing the models with variable gas accretion timescale.  
	}
	\label{gce-tau}
\end{figure*}

The metal-enriched gas accretion scenario introduces an environmental dependence of global gas metallicity by providing an external source of metals that varies with environment. Alternatively, the gas metallicity can vary through changes of the gas content.

The initial gas reservoir and gas accretion rate is fixed to today's observation based on the constant SFH assumption. 
Hence the gas content can be modified in the model through a variation of the gas accretion timescale, i.e.\ the $e$-folding time of the exponential decline (note that the gas stripping process is not taken into account in our phenomenological model).
    
Figure~\ref{gce-tau} shows the prediction of models with different gas accretion timescales in the radial profiles of gas metallicity (left panel) and $\Sigma_{\rm SFR}$ (right panel). Similar to Figure~\ref{gce-enrich}, solid lines of different colours are models with different gas accretion timescales. The dashed lines indicate the initial status at a lookback time of 2 Gyr, which is the same as Figure~\ref{gce-enrich}. The gas accretion timescales in each set of models are assumed to be radially independent. Pristine gas accretion with zero metal content is assumed in these models.

It can be seen that also this scenario is successful at reproducing an increase of gas metallicity. Models with shorter timescales of gas accretion produce higher gas metallicities. This is because a shorter gas accretion timescale minimises the dilution effect. This can then be linked to galaxy environment assuming that accretion timescales are shorter in high densities.

Different from the enriched gas accretion scenario, the radial gradient of gas metallicity and $\Sigma_{\rm SFR}$ changes significantly. A steeper gradient in gas metallicity and a flatter gradient in SFR is found for models with shorter gas accretion timescales. We note, however, that these trends are both the opposite of the observed trends in \textsection3. 

To reverse the trend of the gradient in gas metallicity and $\Sigma_{\rm SFR}$ to match the observations, a radial dependence of the the gas accretion timescale needs to be introduced. 
A shorter gas accretion timescale at large radii will enhance the gas metallicity and decrease $\Sigma_{\rm SFR}$ leading to flatter metallicity gradients and steeper SFR gradients as observed. In the following we will discuss this scenario in more detail.

\subsection{Best-fit model}

\subsubsection{Comparison with data}
As illustrated in the above sections, in principle both scenarios with enriched gas accretion or variable gas accretion timescale are able to introduce the necessary variation in global gas metallicity. But to match the observed trends of the radial gradients in these quantities, we need to invoke additional radial dependence in both scenarios.

\begin{figure*}
	\centering
	\includegraphics[width=18cm]{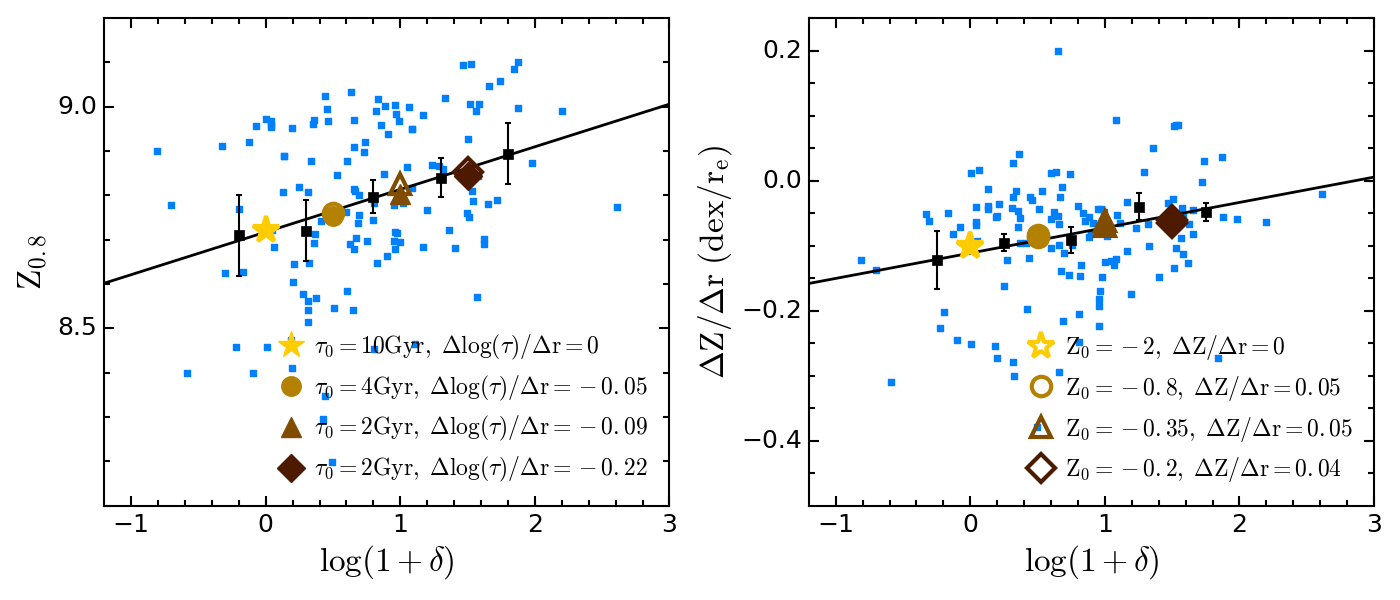}
	\caption{Comparison between the observations and {predictions} of model with enriched gas accretion (empty symbols) and variable gas accretion timescales (filled symbols) in gas metallicity at 0.8$r_{\rm e}$ and its gradient. 
    {Four sets of models are calculated for each scenario. The legend lists the values of the varying parameters in these models. They provide the variable accreted metallicity ${\rm Z}$ at the galactic centre (${\rm Z_0}$) and its radial variation ($\Delta$${\rm Z}/\Delta$r) for the former scenario and the variable central 
	accretion timescale ($\tau_0$) and its variation slope with radius ($\Delta$log($\tau$)$/\Delta$r) for the latter scenario. Blue squares, black errors and solid lines are the same as the left-hand panels in Fig~\ref{absm-obs}, representing the observations in low-mass satellites.}
	}
	\label{enrich-tau-data}
\end{figure*}

Here we perform a direct comparison between the prediction of the models with such radial dependence to the observations. 
Figure~\ref{enrich-tau-data} shows the comparison. Blue data points represent the observations for the low-mass satellites and solid lines represent the best linear fit. These are taken from Figure~\ref{profile-obs}. Large symbols indicate the prediction of the models. Four sets of models are ran for each scenario. Open symbols are the models for the enriched gas accretion scenario while filled symbols are the models for variable gas accretion timescales as shown by the legend. 
{In the variable accretion timescale model, the accretion timescale decreases with galactocentric radius with the central value indicated by $\tau_0$ and the decreasing rate by $\Delta$log($\tau$)$/\Delta$r. In enriched accretion model, the metallicity in gas accretion instead increases with radius with the central value indicated by $[{\rm Z_0}]$ and the increasing rate by $\Delta$$[{\rm Z}]/\Delta$r.}

Both model flavours are successful at reproducing the observed variation of global gas metallicity and its gradient as a function environment.

In the model with variable gas accretion timescales, the gas accretion is required to be shorter in denser environments at at larger galactic radii. For example, in a dense environment with overdensity log(1+$\delta$)=1.5, accretion timescales as short as 0.06 Gyr are required in the outskirts of low-mass satellites (at $1.5 r_{\rm e}$). 

In the alternative enriched gas accretion scenario, the gas accretion to the low-mass satellites residing in denser environments and at large galactic radii has to be more metal-enriched. The metallicity in the gas accreted at $1.5 r_{\rm e}$ in the densest environment with overdensity log(1+$\delta$)=1.5 is extremely high, reaching 0.3 dex (i.e two times above the solar metal abundance). It is worth noting that even the most massive galaxies in the local universe have metallicity below this value (e.g. \citealt{thomas2010}). 


\subsubsection{Gas accretion history}
\begin{figure*}
	\centering
	\includegraphics[width=16cm]{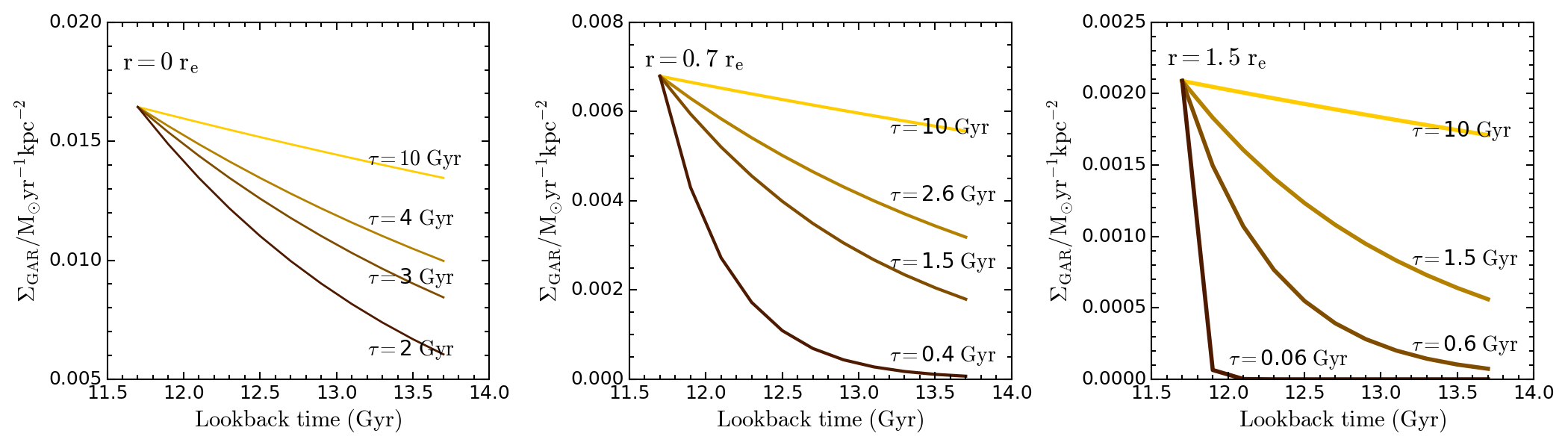}
	\caption{Gas accretion history at three radial bins in each set of models in Figure~\ref{enrich-tau-data} in the past 2 Gyr. The gas accretion timescale of the model with enriched accretion scenario is fixed to be 10 Gyr, the fiducial timescale. Other model tracks represent the gas accretion history of models with variable gas accretion timescale. The timescale for each model is marked in the plot accordingly.        
	}
	\label{gah}
\end{figure*}

We show the gas accretion histories over the past 2 Gyr for these models in Figure~\ref{gah}. For each set of models, the gas accretion histories for three radial bins (0$r_{\rm e}$, $0.7 r_{\rm e}$, $1.5 r_{\rm e}$) are shown in each of the three panels. In the model with enriched gas accretion (light orange lines), the gas accretion timescale is set to be the fiducial timescale of 10 Gyr. Therefore the gas accretion histories of these models are the same. The darker lines show the gas accretion histories of the models for the scenario with variable gas accretion timescales. The timescale for each model is marked in the plot accordingly.

\subsubsection{Radial metallicity profile}
\begin{figure*}
	\centering
	\includegraphics[width=14cm]{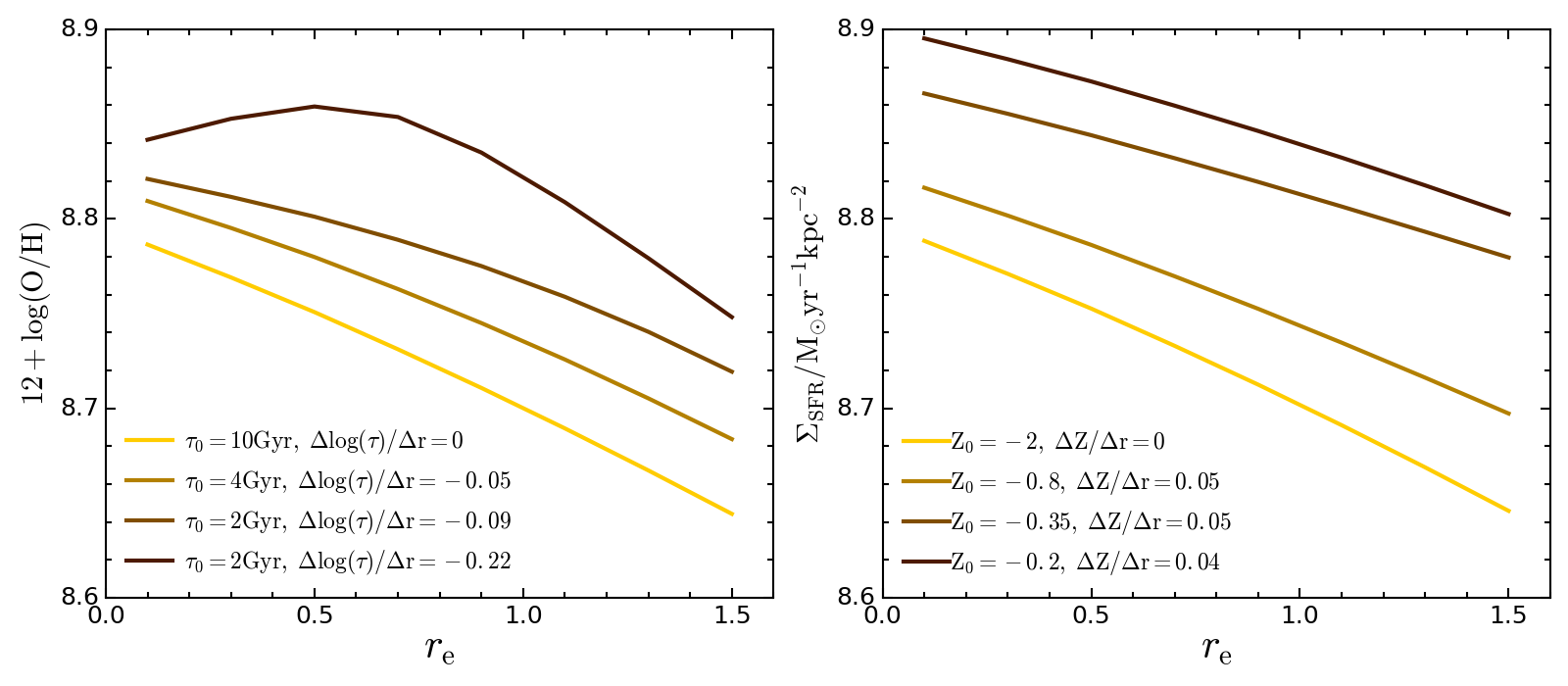}
	\caption{Radial profile of gas metallicity predicted by the model with variable accretion timescale (left-hand panel) or enriched gas accretion (right-hand panel). The adopted {accretion} timescale and metallicity of gas accretion and their radial variation slope are indicated in the legend at the bottom-left corner {which are the same as Fig~\ref{enrich-tau-data}}.    
	}
	\label{enrich-tau-profile}
\end{figure*}

The predicted radial profile of gas metallicity by the best-fit model with either variable gas accretion timescales (left-hand panel) or enriched gas accretion (right-hand panel) are shown in Figure~\ref{enrich-tau-profile}. The observed flatter radial profiles of gas metallicity in denser environments are well replicated by the both models.  
It is interesting to note that the radial profile predicted by the variable gas accretion timescale scenario in dense environment show a similar behaviour to the observations with a flatter gradient in inner regions. 

\section{Discussion}

\subsection{Dynamical friction timescale}
To investigate whether and how dynamical friction timescales (effectively the lookback time at which the model starts) affect our results, we test the model with enriched gas accretion that evolves from a lookback time of 5 Gyr. The initial gas mass and SFR are the same as the previous models. To account for the evolution of gas metallicity in the past 5 Gyr, we assume the gas metallicity at a lookback time of 5 Gyr to be lower than today by 0.5 dex. All other parameters are the same as in the models in Figure~\ref{gce-enrich}. 

Figure~\ref{gce-enrich-5gyr} shows the resulting predictions. It can be seen that the trend introduced by enriched gas accretion remains unchanged. This trend has even become clearer because of the larger evolutionary time elapsed as expected. Therefore adopting a different lookback time for the onset of the evolution will not affect the ability of the scenario to match the observations but may change the exact value of parameters that best fit the data. The same conclusion holds for the variable accretion timescale scenario. 

\subsection{$\Sigma_{\rm SFR}$ with environment}

The star formation rate density $\Sigma_{\rm SFR}$ and its gradient provide additional constraints. In the previous sections we found a weak dependence of $\Sigma_{\rm SFR}$ on the environment with lower global $\Sigma_{\rm SFR}$ and steeper $\Sigma_{\rm SFR}$ gradient in denser environments  \citep[see also][]{schaefer2017,finn2018,spindler2018}. Since the trend of $\Sigma_{\rm SFR}$ with environment is not very significant (at the $\sim 2\sigma$ level), we do not use them to directly constrain our model. However, it is worth noting that the two scenarios discussed here produce different predictions for $\Sigma_{\rm SFR}$ and its gradient. Therefore $\Sigma_{\rm SFR}$ could provide important insight on disentangling these two scenarios.  

In the variable accretion timescale scenario, the trend of global $\Sigma_{\rm SFR}$ is consistent with observations but the gradient is the opposite of the observed trend (i.e. flatter gradient in denser environment, see Figure~\ref{gce-tau}).

However, as discussed in the previous sections, a radial dependence of the gas accretion timescale is required to match the dependence of the gas metallicity gradient on environment. A shorter accretion timescale at large radii is adopted in the best-fit model. This radial dependence is required to be steeper in the denser environment. The shorter accretion timescale shuts down the gas supply and suppresses the star formation at larger radii, resulting in a steeper $\Sigma_{\rm SFR}$ gradient. Therefore, with the radially dependent accretion timescale, the variable accretion timescale scenario can also reproduce the observed trend of global $\Sigma_{\rm SFR}$ and its gradient with environment. Therefore the variable accretion timescale scenario is more favoured by the $\Sigma_{\rm SFR}$ observations than the enriched gas accretion scenario, if the trend of $\Sigma_{\rm SFR}$ with environment is valid.    

In the enriched gas accretion scenario, since the star formation history is not affected, no variation of $\Sigma_{\rm SFR}$ and its gradient is predicted for different environments. This puts the scenario in tension with the observed dependence of $\Sigma_{\rm SFR}$ on environment.

Moreover this scenario requires the gas accreted at larger radii to be more metal-enriched than at small radii. This does not seem physically plausible. In particular, this scenario requires the metallicity of the accreted gas to be as high as twice solar, which seems very high and is not supported by observations of the circumgalactic medium \citep{prochaska2017}. Therefore we consider the enriched gas accretion scenario to be a less likely explanation for the observed dependence of gas metallicity and its gradient on environment.  



\subsection{Outside-in quenching}

As implied by our successful model with variable gas accretion timescales, the gas accretion onto low-mass satellites is preferentially suppressed in dense regions, and preferentially at large galactic radii. This is consistent with the `environment quenching' picture proposed by \citet{peng2010} where the cessation of star formation is believed to be more efficient in denser regions \citep{peng2012}.   

The additional radial dependence found in this work suggests an outside-in suppression of star formation as that we refer to as `outside-in quenching'. This picture was used before to explain the more centrally concentrated star formation in galaxies in denser environments in the literature \citep{schaefer2017,finn2018,spindler2018}. An outside-in mass assembly mode in low-mass galaxies was also proposed by \citet{pan2015} to explain the negative or flat ${\rm NUV}-r$ colour gradients in these galaxies. To further explore the physical processes that are responsible for the outside-in quenching or radially dependent suppression of gas accretion, hydrodynamical simulations with high resolution will be necessary which is beyond the scope of this work. 






\section{Summary}

We investigate the environmental dependence of gas metallicity and star formation rate surface density ($\Sigma_{\rm SFR}$) and their radial distribution in local star-forming galaxies based on SDSS/MaNGA IFU survey data. We select a large galaxy sample of 1414 galaxies which allows us to split our sample into different mass bins to minimize the mass dependence. 
To quantify the galaxy environment, we first split our galaxy sample into central and satellite galaxies and then use the number density around each galaxy to indicate their local environment. 

We first explore the environmental dependence in a qualitative way by inspecting the radial profiles of gas metallicity and $\Sigma_{\rm SFR}$ in four environment bins. 
We find no significant dependence of gas metallicity on galaxy environment except for low-mass satellites with $M_*<10^{9.7}{\rm M_{\odot}}$. For these, the gas metallicity tends to be systematically higher and its negative metallicity gradient systematically steeper when the satellites reside in denser environments. 
The dependence of $\Sigma_{\rm SFR}$ on environmental density, instead, is generally weak. 
 
We then study the environmental dependence in a quantitative way. We measure the global value and the gradient of gas metallicity and $\Sigma_{\rm SFR}$ for each galaxy and assess their correlation with environmental density. The qualitative inspection above and the quantitative analysis give consistent results on the gas metallicity. For $\Sigma_{\rm SFR}$ we identify a weak dependence of $\Sigma_{\rm SFR}$ and its radial gradient on environment. The $\Sigma_{\rm SFR}$ of satellites residing in denser regions is generally lower and the radial gradient steeper.  

To understand the physical origin of the environmental dependence of gas metallicity and $\Sigma_{\rm SFR}$, we use a chemical evolution model to reproduce the observed trends. Interestingly, we find two scenarios that are able to explain the dependence of global gas metallicity and its gradient on environment equally well. One scenario assumes metal-enriched gas accretion. In this scenario gas with higher metallicity is accreted both in denser environments and at larger radii. The other scenario assumes variable gas accretion timescales with shorter accretion timescales in denser environments and at larger radii. 

It turns out that star formation rate density helps to break the degeneracy between these two scenarios. Different from the gas metallicity, the observed dependence of $\Sigma_{\rm SFR}$ on environment favours the scenario with variable accretion timescales. Because of the shorter accretion timescales at large radii within this scenario, this solution leads to an outside-in quenching picture for low-mass satellites.

Note that the environmental dependence of the gas metallicity and $\Sigma_{\rm SFR}$ gradients are significant at a 2$-4\sigma$ level. The metal-enriched gas accretion scenario cannot be ruled out entirely. A larger galaxy sample with high quality IFU observations is needed in the future to assess the environmental dependence of galaxy properties with a better statistic and to pin down the underlying physical origin.

\section*{Acknowledgements}
We are very grateful to the referee, Rolf Kudritzki, for the insightful report that helped improve the paper considerably.

The Science, Technology and Facilities Council is acknowledged for support through the Consolidated Grant ‘Cosmology and Astrophysics at Portsmouth’, ST/N000668/1. Numerical computations were done on the Sciama High Performance Compute (HPC) cluster which is supported by the ICG, SEPnet and the University of Portsmouth.

Funding for the Sloan Digital Sky Survey IV has been provided by the Alfred P. Sloan Foundation, the U.S. Department of Energy Office of Science, and the Participating Institutions. SDSS acknowledges support and resources from the Center for High-Performance Computing at the University of Utah. The SDSS web site is www.sdss.org.

SDSS is managed by the Astrophysical Research Consortium for the Participating Institutions of the SDSS Collaboration including the Brazilian Participation Group, the Carnegie Institution for Science, Carnegie Mellon University, the Chilean Participation Group, the French Participation Group, Harvard-Smithsonian Center for Astrophysics, Instituto de Astrofísica de Canarias, The Johns Hopkins University, Kavli Institute for the Physics and Mathematics of the Universe (IPMU) / University of Tokyo, the Korean Participation Group, Lawrence Berkeley National Laboratory, Leibniz Institut für Astrophysik Potsdam (AIP), Max-Planck-Institut für Astronomie (MPIA Heidelberg), Max-Planck-Institut für Astrophysik (MPA Garching), Max-Planck-Institut für Extraterrestrische Physik (MPE), National Astronomical Observatories of China, New Mexico State University, New York University, University of Notre Dame, Observatório Nacional / MCTI, The Ohio State University, Pennsylvania State University, Shanghai Astronomical Observatory, United Kingdom Participation Group, Universidad Nacional Autónoma de México, University of Arizona, University of Colorado Boulder, University of Oxford, University of Portsmouth, University of Utah, University of Virginia, University of Washington, University of Wisconsin, Vanderbilt University, and Yale University.

\appendix
\section{Density by $N-$body simulation}

\begin{figure*}
	\centering
	\includegraphics[width=16cm]{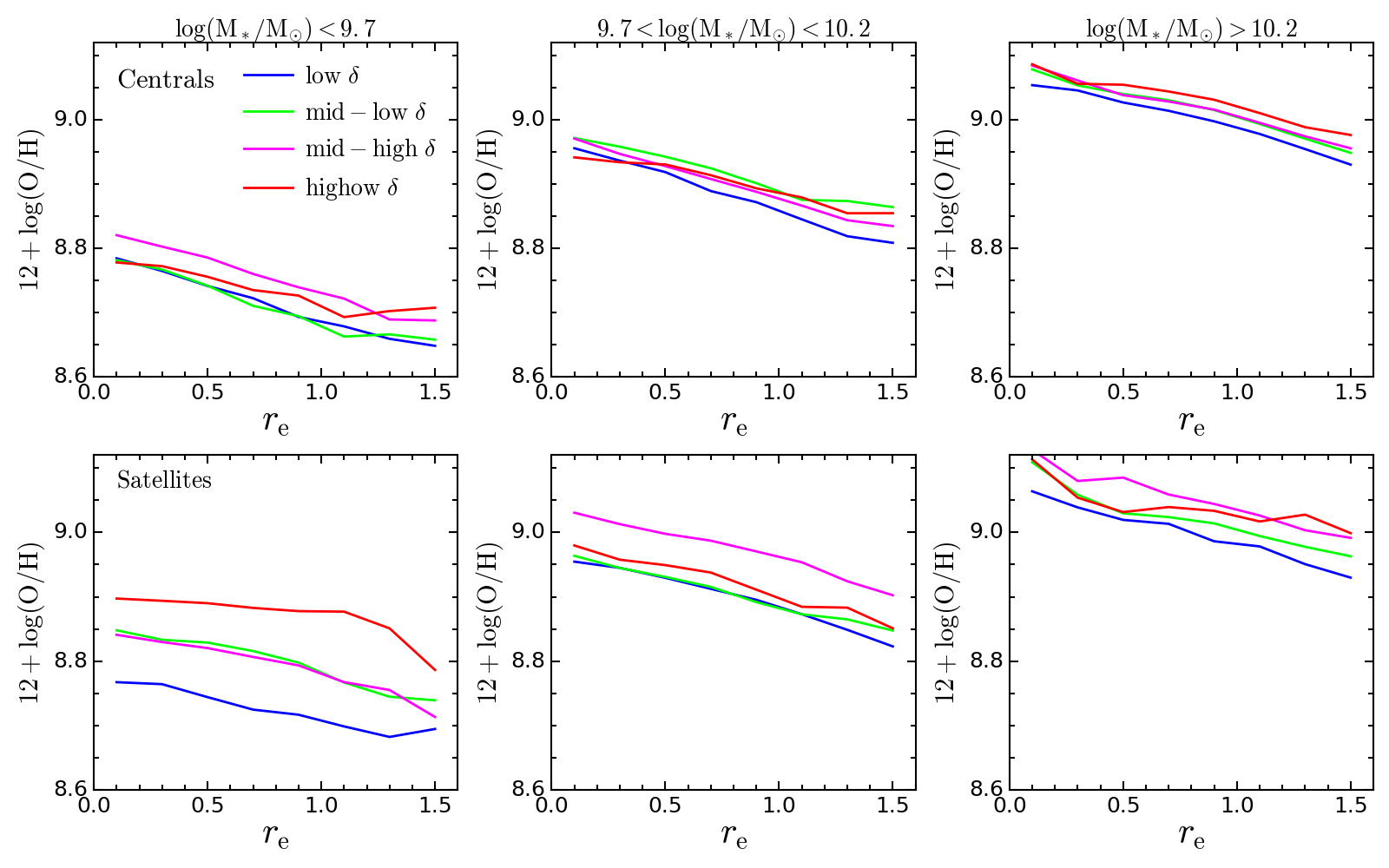}
	\caption{The same as Figure ~\ref{profile-obs} except that the environmental densities are obtained within 3Mpc in the density field reconstructed by $N$-body simulations.    
	}
	\label{profile-lss}
\end{figure*} 

\begin{figure*}
	\centering
	\includegraphics[width=16cm]{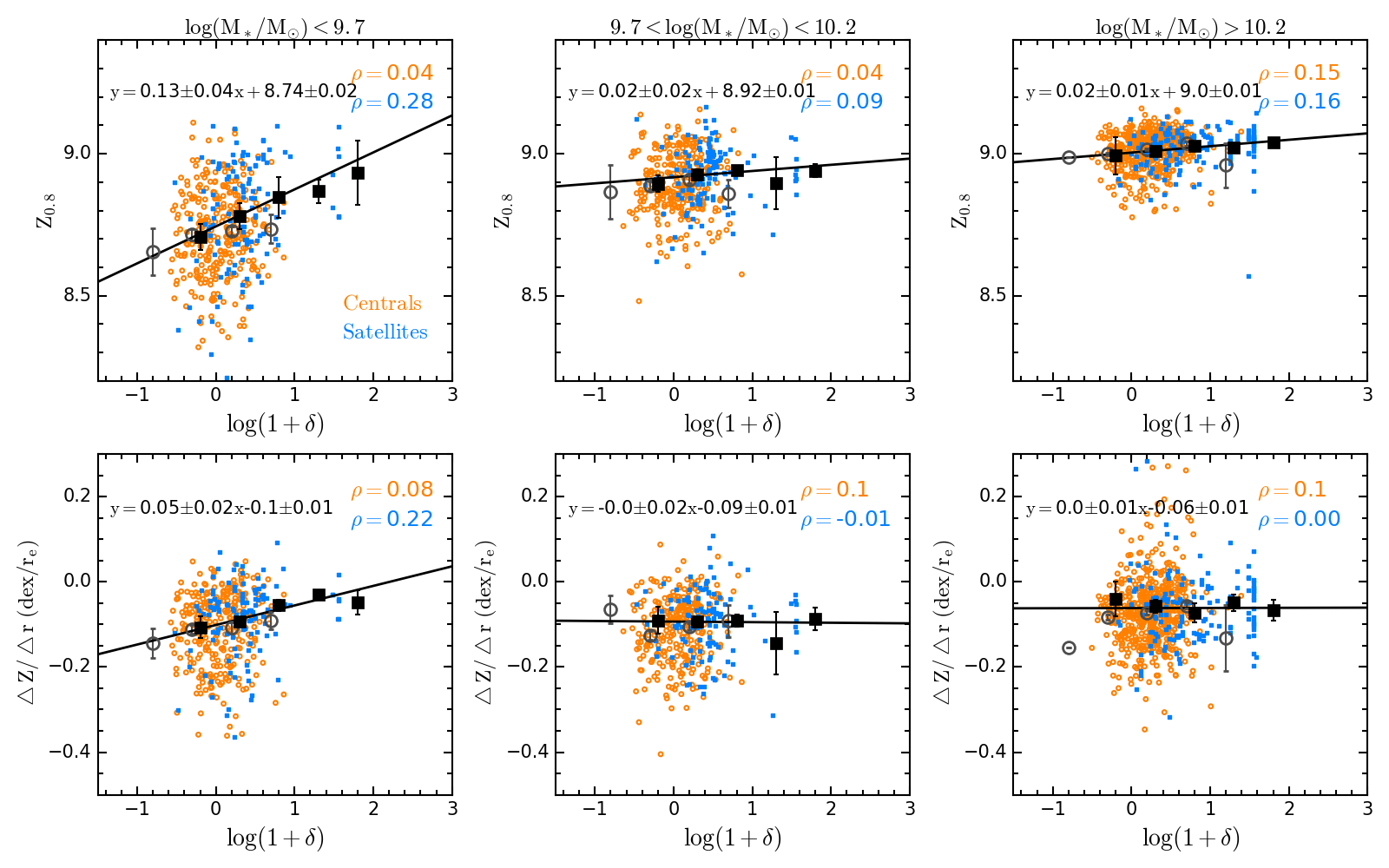}
	\caption{The same as Figure ~\ref{absm-obs} except that the environmental densities are obtained within 3Mpc in the density field reconstructed by $N$-body simulations.    
	}
	\label{absm-lss}
\end{figure*} 

\section{Start evolution at 5 Gyr ago}

\begin{figure*}
	\centering
	\includegraphics[width=14cm]{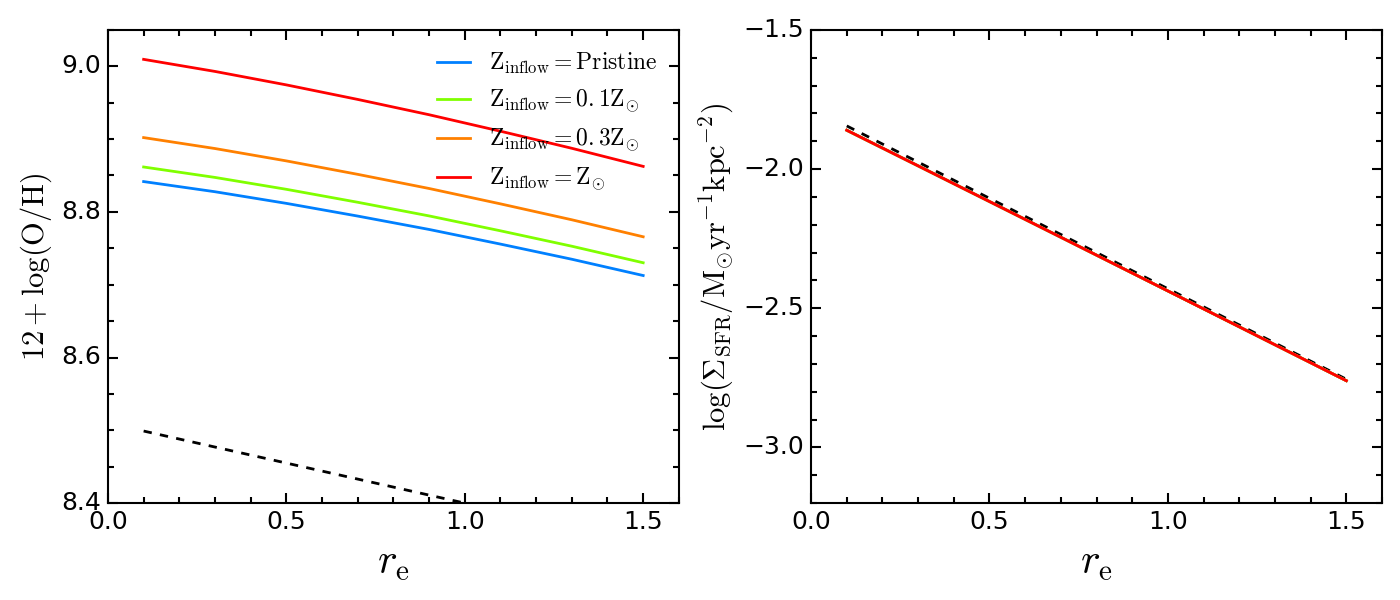}
	\caption{The same as Figure~\ref{gce-enrich} but the model starts evolve from a lookback time of 5 Gyr instead of 2 Gyr ago. The initial gas metallicity is assumed to 0.5 dex instead of 0.2 dex lower than today's observation.  
	}
	\label{gce-enrich-5gyr}
\end{figure*}
\end{document}